\documentclass[12pt]{article}
\usepackage{hyperref}
\newtheorem{theorem}{Theorem}
\newtheorem{proposition}{Proposition}
\newtheorem{lemma}{Lemma}

\newtheorem{definition}{Definition}
\newtheorem{example}{Example}
\newenvironment{proof}[1][Proof]{\begin{trivlist}
\item[\hskip \labelsep {\bfseries #1}]}{\qed\end{trivlist}}
\newcommand{\qed}{\hfill\vrule height 1ex depth 0pt width 1ex}
\def\g{\gamma}
\def\l{\lambda}
\def\D{{\cal D}}
\def\U{{\cal U}}
\def\implies{\to}
\def\st{\bigm|}
\title{Quantitative Redundancy \\ in Partial Implications}
\author{Jos\'e L. Balc\'azar%
\thanks{Partially supported by 
project BASMATI (TIN2011-27479-C04-04) of Programa Nacional de 
Investigaci\'on (Ministerio de Ciencia e Innovaci\'on, Spain) and
grant 2014SGR 890 (MACDA) from AGAUR, Generalitat de Catalunya.} \\
Department of Computer Science \\
Universitat Polit\`ecnica de Catalunya \\
\url{jose.luis.balcazar@upc.edu}}

\begin{document}

\maketitle

\begin{abstract}
We survey the different properties of an 
intuitive notion of redundancy, as a function 
of the precise semantics given to the notion
of partial implication.

The final version of this survey will appear
in the Proceedings of the Int.~Conf.~Formal
Concept Analysis, 2015.
\end{abstract}

\section{Introduction}

The discovery of regularities in large scale data is a multifaceted 
current challenge. Each syntactic mechanism proposed to represent
such regularities opens the door to wide research questions.  
We focus on a specific sort of regularities sometimes found in 
transactional data, that is,
data where each observation is a set of items, and defined 
in terms of pairs of sets of items.

Syntactically, the fact that this sort of regularity holds for a given 
pair $(X,Y)$ of sets of items is often denoted as an implication: 
$X \to Y$. 
However, whereas in Logic an implication like this is true if and only 
if $Y$ holds whenever $X$ does, in our context, namely, partial 
implications and association rules, it is enough if $Y$ holds 
``most of the times'' $X$ does. Thus, in association mining, 
the aim is to find out which expressions of that sort are valid
for a given transactional dataset: for what $X$ and what $Y$, 
the transactions that contain $X$ ``tend to contain'' $Y$~as~well. 

In many current works, that syntax is defined as if its meaning
was sufficiently clear. Then, any of a number of ``measures of 
interestingness'' is chosen to apply to them, in order to select some
to be output by a data analysis process on a particular dataset.
Actually, the mere notation $X \to Y$ is utterly insufficient: 
any useful perspective requires 
to endow these expressions with a definite semantics that makes 
precise how that na\"\i{}ve intuition of ``most of the times'' 
is formalized; only then can we study and clarify the algorithmic 
properties of these syntactical expressions. Thus, we are not 
really to
``choose a measure of interestingness'' but plainly to \emph{define} 
what $X\to Y$ means, and there are many acceptable ways of doing this. 

This idea of a relaxed implication connective is a relatively 
natural concept, and versions sensibly defined by resorting to
conditional probability have been proposed in different
research communities: a common semantics of $X\to Y$
is through a lower bound on its ``confidence'', the conditional probability 
of $Y$ given~$X$. This meaning appears already in the ``partial 
implications'' of \cite{Lux} (actually, ``implications partielles'', 
with confidence christened there ``pr\`ecision'').
Some contributions based on 
Mathematical Logic develop notions 
related to these partial implications defined in terms of conditional 
probability: see 
\cite{GUHA}. 
However, it must be acknowledged that the 
contribution that turned on the spotlights on partial implications
was \cite{AgrawalImielinskiSwami} and
the improved algorithm in \cite{AMSTV}:~the proposal of exploring
large datasets in search for association rules of high support and
confidence has led to huge amounts of research since.
Association rules are
partial implications that impose the additional condition that 
the consequent is a single item.

Three of the major foci of research in association rules and
partial implications are as 
follows. First, the quantity of candidate itemsets for both the
antecedent $X$ and, sometimes, the consequent $Y$ grows 
exponentially with the 
number of items. Hence, the space to explore is potentially enormous:
on real world data,
very soon
we run already into billions of candidate antecedents.
Most existing solutions are based on the acceptance that, as 
not all of them 
can be considered within 
reasonable running times, we make do with those that obey the 
support constraint (``frequent itemsets''). The support constraint
combines well with confidence in order to avoid reporting mere
statistical artifacts \cite{MegSrik} but its major role
is to reduce the search space. A wide repertory of algorithms for 
frequent sets and association rule mining exists by now 
\cite{FqPattBook}.

Second, many variations have been explored: for instance, cases of more 
complicated structures in the data 
and, also, combinations with other machine-learning models 
or tasks like in \cite{JarSchSim,YinHan}. 

This paper surveys part of a research line that belongs to a third focus:
in a vast majority of practical applications, if any partial
implication is found at all, it often happens that the search 
returns hundreds of thousands of them. 
It is far from trivial to design an associator able to choose 
well, among them, a handful to show to an impatient user. This is
tantamount to modifying the semantics of the partial implication
connective, by adding or changing the conditions under which one
such expression is deemed valid and is to be reported. Most often,
but not always (as we report in Sections 
\ref{s:reprules}~and~\ref{s:multprem})
this approach takes the form of ``quality evaluations'' performed 
to select which partial implications are to be highlighted for 
the user. We do not consider this problem solved yet, but deep
progresses have been achieved so far; we survey a humble handful
of those, where the present author was actively involved.
For a wider perspective of all these three aspects of association 
rule mining, see Part II of~\cite{ZakiMeira}.

The main link along this paper can be described informally
as follows: human intuition, maybe on the basis of our experience 
with full, standard implications, tends to expect that smaller
antecedents are better than larger ones, and larger consequents
are better than smaller ones. We call this statement here the
\emph{central intuition} of this paper; many references express,
in various variants, this intuition
(e.g.~\cite{BAG,KryszPKDD,LiuHsuMa,PadTu2000,ShahLaksRS,ToKleRHM}  
just to name a few). This intuition is only partially true in
implications, where the GD basis gets to be minimal through the
use of subtly enlarged antecedents~\cite{GD}. This survey paper discusses,
essentially, the particular fact that, on partial implications, 
this intuition is both true and false\dots\ as a function, of course, 
of the actual semantics given to the partial implication connective.

\section{Notation and Preliminary Definitions}


Our datasets are transactional. This means 
that they are composed of transactions, each of which consists
of an itemset with a unique transaction identifier. Itemsets
are simply subsets of some fixed set $\U$ of items.
We will denote itemsets by capital letters from the end of the
alphabet, and use juxtaposition to denote union, as in $XY$. 
The inclusion sign as in $X\subset Y$ denotes proper subset,
whereas improper inclusion is denoted $X\subseteq Y$.
The cardinality of a set $X$ (either an itemset or a set of
transactions) is denoted $|X|$.

\subsection{Partial Implications}

As indicated in the Introduction, the most common semantics of 
partial implication is its \emph{confidence}: the conditional
empirical probability of the consequent given the antecedent,
that is, the ratio between the number of transactions in which 
$X$ and $Y$ are seen together and the number of transactions 
that contain~$X$. We will see below that this semantics may be 
somewhat misleading. In most application cases, the search 
space is additionally restricted by a minimal \emph{support} 
criterion, thus avoiding itemsets that appear very seldom 
in the dataset.
 
More precisely, for a given dataset~$\D$, consisting of~$n$ 
transactions, 
the \emph{supporting set} $\D_X\subseteq\D$ 
of an itemset $X$ is the subset of transactions that include $X$.
(For the reader familiar with the FP-growth frequent set miner
\cite{HPYM04},
these are the same as their ``projected databases'', except for
the minor detail that, here, we do not remove $X$ from the 
transactions.)

The \emph{support} $s_{\D}(X) =  |\D_X|/n \in[0,1]$ of an itemset $X$
is the cardinality of the set of transactions that contain~$X$ divided 
by~$n$; it corresponds to the relative frequency or empirical 
probability of $X$. An alternative rendering of support is its 
unnormalized version, but some of the notions that will play a 
major role later on are simpler to handle with normalized supports. 
Now, the \emph{confidence} of a partial implication $X\to Y$ 
is $c_{\D}(X\to Y) = s_{\D}(XY)/s_{\D}(X)$:
that is, the empirical approximation to the corresponding conditional 
probability. The \emph{support} of a partial implication $X\to Y$ is 
$s_{\D}(X\to Y) = s_{\D}(XY)$. In both expressions, we will omit
the subscript $\D$ whenever the dataset is clear from the context.
Clearly, 
$s_{\D_Z}(X) = \frac{|\D_{XZ}|}{|\D_Z|} = c(Z\to X)$.

Often, we will assume that $X\cap Y = \emptyset$ in partial 
implications $X\to Y$. Some works impose this condition globally;
we will mention it explicitly whenever it is relevant, but,
generally speaking, we allow $X$ and $Y$ to intersect or, even,
to fulfill $X\subseteq Y$. 
Note that, if only support and confidence are at play, then
$c_{\D}(X\to XY) = c_{\D}(X\to Y)$ and 
$s_{\D}(X\to XY) = s_{\D}(X\to Y)$.
Of course, in practical terms, after
a partial implication mining process, only the part of $Y$ that
does not appear in $X$ would be shown to the user.

We do allow $X=\emptyset$ as antecedent of a partial implication:
then, its confidence coincides with the support,
$c_{\D}(\emptyset\to Y) = s_{\D}(Y)$, since $s_{\D}(\emptyset)=1$.
Allowing \hbox{$Y=\emptyset$} as consequent as well is possible
but turns out not to be very useful; therefore,
empty-consequent partial implications are always omitted from consideration.
All along the paper, there are occassional glitches where the
empty set needs to require separate consideration. Being interested
in the general picture, here we will mostly ignore these issues,
but the reader can check that these cases are given careful treatment
in the original references provided for each part of our discussion.

By $X \Rightarrow Y$ we denote full, standard logical 
implication; this expression will be called
the \emph{full counterpart} of the partial implication 
$X\to Y$. 

\subsection{Partial Implications versus Association Rules}

Association rules were defined originally as partial implications 
$X\to Y$ with singleton consequents: $|Y| = 1$; we abbreviate
$X\to \{A\}$ as $X\to A$. This decision allows one to reduce 
association mining to a simple 
postprocessing after finding frequent sets. Due to the illusion of 
augmentation, many users are satisfied with this syntax, but,
however, more items in the consequent provide more information.

Indeed, in full implications, 
the expression $(A\Rightarrow B)\land(A\Rightarrow C)$
is fully equivalent to $A\Rightarrow BC$, and we lose little by
enforcing singleton consequents (equivalently, definite Horn clauses);
an exception is the discussion of minimal bases, where nonsingleton
consequents allow for canonical bases that are unreachable in the
Horn clause syntax~\cite{GD}. But, in partial implications, $A\to BC$ says
more than the conjunction of $A\to B$ and $A\to C$,
namely, $B$ and $C$ abound \emph{jointly} in $\D_{A}$. Whenever
possible, $A\to BC$ is better, being both more economical and
more informative. This can be ilustrated by the following example
from~\cite{BalTKDD}, to which we will return later on.

\begin{example}
\label{ex:running}
Consider a dataset on $\U = \{ A, B, C, D, E \}$ consisting of
12 transactions: 6 of them include all of $\U$, 2 consist of
$ABC$, 2 more are $AB$, and then one each of $CDE$ and $BC$.
It can be seen that the confidence of both $B\to A$ and $B\to C$
is $9/11$, whereas the confidence of $B\to AC$ is $8/11$. 
\end{example}

Actually, even restricted to association rules, the output of
confidence-based associators is often still too large: the rest of
this paper discusses how to reduce the output
with no loss of information, first, and, then, as the outcome is 
often still too large in practice, we will need to allow for
a carefully tuned loss of information.

\section{Redundancy in Confidence-Based Partial Implications}\label{s:reprules}

We start our discussion by ``proving correct'' 
our \emph{central intuition},
that is, providing a natural semantics under which that intuition 
is correct. For this section, we work under confidence and support 
thresholds, and it turns out to be convenient to explicitly 
assume that the left-hand side of each partial implication 
is included in the right-hand side. We force that inclusion 
using notations in the style of $X\to XY$.

Several references (\cite{AgYu} for one) 
have considered the following argument:
assume that we could know beforehand that, in all datasets,
the confidence and support of $X_0\to X_0Y_0$ 
are always larger than or equal to those of $X_1\to X_1Y_1$.
Then, whenever we are mining some dataset under confidence and
support thresholds, assume that we find $X_1\to X_1Y_1$: 
we should not bother to report as well $X_0\to X_0Y_0$, since 
it must be there anyhow, and its presence in the output is
uninformative. In a very strong sense, $X_0\to X_0Y_0$ is
redundant with respect to $X_1\to X_1Y_1$. Irredundant 
partial implications
according to this criterion are called ``essential rules''
in \cite{AgYu} and \emph{representative rules} in
\cite{KryszPAKDD}; we will follow this last term.


\begin{lemma}
\label{l:redchar}
Consider two partial implications, $X_0\to X_0Y_0$ 
and $X_1\to X_1Y_1$.
The following are equivalent:
\begin{enumerate}
\item 
The confidence and support of $X_0\to X_0Y_0$ 
are larger than or equal to those of $X_1\to X_1Y_1$, 
in {\em all} datasets: for every $\D$, 
$c_{\D}(X_0\to X_0Y_0)\geq c_{\D}(X_1\to X_1Y_1)$ and 
$s_{\D}(X_0\to X_0Y_0)\geq s_{\D}(X_1\to X_1Y_1)$.
\item 
The confidence of $X_0\to X_0Y_0$ 
is larger than or equal to that of $X_1\to X_1Y_1$, 
in {\em all} datasets: for every $\D$, 
$c_{\D}(X_0\to X_0Y_0)\geq c_{\D}(X_1\to X_1Y_1)$.
\item $X_1\subseteq X_0\subseteq X_0Y_0\subseteq X_1Y_1$.
\end{enumerate}
\end{lemma}

When these cases hold, we say that 
$X_1\to X_1Y_1$ makes $X_0\to X_0Y_0$ 
\emph{redundant}. The fact that the inequality on
support follows from the inequality on confidence
is particularly striking. This lemma can be interpreted
as proving correct the \emph{central intuition} 
that smaller antecedents
and larger consequents are better, by indentifying a
semantics of the partial implication connective that makes
this true and by pointing out that it is not just the
consequent that is to be maximized, but the union of
antecedent and consequent. If only consequents are 
maximized separately, and are kept disjoint from the antecedents, 
then one
gets to a quite more complicated situation discussed below.


\begin{definition}
Fix a dataset and confidence and support thresholds. 
The \emph{representative rule basis} 
for that dataset at these support
and confidence thresholds
consists of
those partial implications that pass both thresholds in the dataset,
and are not made redundant, 
in the sense of the previous paragraph, 
by other partial implications also above the 
thresholds. 
\end{definition}


Hence, a redundant partial implication is so because we can know 
beforehand, from the information in the basis, that its 
confidence is above the threshold. We have:

\begin{proposition}
(Essentially, from \cite{KryszPAKDD}.)
For a fixed dataset $\D$ and a fixed confidence threshold~$\gamma$:
\begin{enumerate}
\item
Every partial implication of confidence at least $\gamma$
is made redundant by some representative rule.
\item
Partial implication $X\to Y$ with $X\subseteq Y$ is a 
representative rule if and only if 
$c_{\D}(X\to Y) \geq\gamma$ but there is no $X'$ and $Y'$
with $X'\subseteq X$ and 
$XY\subseteq X'Y'$ such that $c_{\D}(X'\to Y') \geq\gamma$,
except $X=X'$ and $Y=Y'$.
\end{enumerate}
\end{proposition}

According to statement \emph{(3)} in Lemma~\ref{l:redchar},
that last point means that
a representative rule is not redundant with respect to
any partial implication (different from itself) that has 
confidence at least $\gamma$ in the dataset. 
It is interesting to note that one does not need
to mention support in this last proposition, the
reason being, of course, statement 
\emph{(2)} in Lemma~\ref{l:redchar}.
The fact that statement \emph{(3)} implies statement \emph{(1)}
was already pointed out in \cite{AgYu,KryszPAKDD,PhanLuongICDM} 
(in somewhat different terms).
The remaining implications
are from~\cite{BalLMCS}; see this reference 
as well for proofs of additional properties,
including the fact the representative basis has the 
minimum possible size among all bases for this notion
of redundancy, and for discussions of
other related redundancy notions.
In particular, several other natural proposals
are shown there to be equivalent to this redundancy.
Also \cite{BalTKDD} provides further properties
of the representative rules. These references discuss
as well the connection with a similar notion in \cite{Zaki}.

In Example~\ref{ex:running}, at confidence threshold 0.8, the
representative rule basis consists of seven partial implications:
$\emptyset\to C$,
$B\to C$,
$\emptyset\to AB$,
$C\to AB$,
$A\to BC$,
$D\to ABCE$, and
$E\to ABCD$.

\subsection{Quantitative Evaluation of Non-Redundancy: 
Confidence Width}
\label{ss:cwidth}

Redundancy is a qualitative property; still, it allows for
a quantitative discussion.
Consider a representative rule $X\to XY$: at confidence
$c(X\to XY)$, no partial implication makes it redundant.
But we could consider now to what extent we need to
reduce the confidence threshold in order to find a partial
implication that would make this one redundant. If a partial
implication of almost the same confidence can be found to 
make $X\to XY$ redundant, then our partial implication 
is not so interesting.
According to this idea, one can define a parameter, the
\emph{confidence width} \cite{Bal09}, that, in a sense, evaluates how
different is our partial implication from other similar ones.
We do not discuss this parameter further, but a related quantity
is treated below in Section~\ref{ss:cboost}.

\subsection{Closure-Aware Redundancy Notions}\label{ss:bstar}

Redundancy of one partial implication with respect to another
can be redefined as well in a similar but slightly
more sophisticate form by taking into account the closure 
operator obtained from the data (see \cite{GanWil99}). Often, 
this variant yields a more economical basis because the full 
implications are described
by their often very short Guigues-Duquenne basis~\cite{GD}; see
again \cite{BalLMCS} for the details.

\section{Redundancy with Multiple Premises}\label{s:multprem}

The previous section indicates precisely when ``one partial
implication follows logically from another''. It is natural
to ask whether a stronger, more useful notion to reduce the
size of a set of partial implications could be based on partial
implications following logically from several others together,
beyond the single-premise case.

Simply considering standard examples with full implications like
Augmentation (from $X\Rightarrow Y$ and $X'\Rightarrow Y'$
it follows $XX'\Rightarrow YY'$) or Transitivity 
(from $X\Rightarrow Y$ and $Y\Rightarrow Z$ it follows 
$X\Rightarrow Z$), it is easy to see that these cases fail
badly for partial implications. Indeed, one might suspect, 
as this author did for quite some time, that one partial 
implication would not follow logically from several premises 
unless it follows from one of them.

Generally speaking, however, this suspicion is wrong. It is
indeed true for confidence thresholds $\g\in(0,0.5)$, but 
these are not very useful in practice, as an association
rule $X\to A$ of confidence less than 0.5 means that, in
$\D_X$, the absence of $A$ is more frequent than its presence.

And, for $\g\in[0.5,1)$, it turns out that, for instance, from 
$A\to BC$ and $A\to BD$ it follows $ACD\to B$, in the sense that
if both premises have confidence at
least $\g$ in any dataset, then the conclusion also does.
The general case for two premises was fully characterized 
in \cite{BalLMCS},
but the case of arbitrary premise sets has remained elusive for 
some years. Eventually, a very recent result from~\cite{AtsBal}
proved that redundancy with respect to a set of premises that
are partial implications hinges on a complicated combinatorial
property of the premises themselves. We give that property a short
(if admittedly uninformative) name here:
\looseness=1

\begin{definition}
\label{df:nice}
Let $X_1 \to Y_1,\ldots,X_k \to Y_k$ be a set of partial implications.
We say that it is \emph{nice} if
$X_1 \Rightarrow Y_1,\ldots,X_k \Rightarrow Y_k \models
X_i \Rightarrow U$, for all $i \in 1\ldots k$, 
where $U = X_1Y_1 \cdots X_kY_k$.
\end{definition}

Here we use the standard symbol $\models$ for logical entailment;
that is, whenever the implications at the left-hand side are true,
the one at the right-hand side must be as well. 


Note that the definition of nicety of a set of partial implications
states a property, not of the partial implications themselves, 
but of their full counterparts.
Then, we can characterize entailment among partial implications
for high enough thresholds of confidence, as follows:

\begin{theorem}
\label{th:mainHGnew}
\cite{AtsBal}
Let $X_1 \to Y_1,\ldots,X_k \to Y_k$ be a set of 
partial implications with $k \geq 1$,
candidates to premises, and a candidate conclusion $X_0 \to Y_0$. 
If $\gamma \geq (k-1)/k$, 
then the following are equivalent:
\begin{enumerate} 
\item in any dataset where the confidence of the premises
$X_1 \to Y_1,\ldots,X_k \to Y_k$ is at least $\gamma$, 
$c(X_0 \to Y_0)\geq\g$ as well;
\item either $Y_0 \subseteq X_0$, or there is a non-empty 
$L \subseteq \{1\ldots k\}$ such that the following conditions hold:
\begin{enumerate} 
\item $\{ X_i \to Y_i : i \in L \}$ is nice,
\item $\bigcup_{i \in L} X_i \subseteq X_0 \subseteq \bigcup_{i \in L} X_iY_i$,
\item $Y_0 \subseteq X_0 \cup \bigcap_{i \in L} Y_i$.
\end{enumerate}
\end{enumerate}
\end{theorem}

Interestingly, the last couple of conditions are reasonably
correlated, for the case of several premises, with the 
\emph{central intuition} that smaller antecedents are better 
than larger ones, and larger consequents are better than 
smaller ones. The premises actually necessary must all 
include the consequent of the conclusion, and their
antecedents are to be included in the antecedent of the
conclusion. Even the additional fact that the antecedent of 
the conclusion does not have ``extra items'' not present
in the premises also makes sense.

However, there is the additional condition that
only nice sets of partial implications may have
a nontrivial logical consequence, 
and all this just for
high enough confidence thresholds. The proof is complex
and we refrain from discussing it here; see \cite{AtsBal},
where, additionally, the case of $\gamma < 1/k$ is also
characterized and the pretty complicated picture for
intermediate values of $\gamma$ is discussed.

We do indicate,
though, that the notion of ``nicety'', in practice, turns out
to be so restrictive that we have not found any case of 
nontrivial entailment from more than one premise in a number
of tests with stardard benchmark datasets. Therefore, this
approach is not particularly useful in practice to reduce
the size of the outcome of an associator.

\subsection{Ongoing Developments}

As for representative rules (Subsection~\ref{ss:bstar}), 
there exists a natural variant of 
the question of redundancy, whereby full implications are handled
separately; essentially, the redundancy notion becomes 
``closure-based''. 
This extension was 
fully 
characterized
as well for the case
of two premises in \cite{BalLMCS}, but it is current work in
progress how to extend the scheme to the case of arbitrary
quantities of premises.

\section{Alternative Evaluation Measures}

We move on to discuss how to reinterpret the \emph{central intuition}
as we change the semantics of the partial implication connective.
Confidence is widely used as a definition of partial implication but, 
in practice, presents two drawbacks. First, it does not detect 
negative correlations; and, second, as already indicated, often 
lets pass far too many rules and, moreover, fiddling with the 
confidence threshold turns out to be a mediocre or just useless 
solution. Examples of both disadvantages are both easy to 
construct and easy to find on popular benchmark datasets. Both objections 
can be addressed by changing the semantics of the expression $X\to Y$, 
by either replacing the confidence measure or by strengthening it
with extra conditions. 
The literature on this topic is huge and cannot be reviewed 
here: see \cite{GH,LencaEtAl,TKS} 
and their references 
for information about the relevant 
developments published along these issues. 
We focus here on just a tiny subset of all these studies.

The first objection alluded to in the previous paragraph
can be naturally solved via an extra 
normalization (more precisely, dividing the confidence by the support 
of the consequent).
The outcome is \emph{lift}, a 
well-known expression in basic probability;
a closely related parameter is \emph{leverage}:

\begin{definition}
\label{df:liftleverage}
Assume $X\cap Y = \emptyset$.
The {\em lift} of partial implication $X\to Y$ is 
$\ell_{\D}(X\to Y) = \frac{c_{\D}(X\to Y)}{s_{\D}(Y)} =
\frac{s_{\D}(XY)}{s_{\D}(X)\times s_{\D}(Y)}$.
The {\em leverage} 
of partial implication $X\to Y$ is 
$\l_{\D}(X\to Y) = s_{\D}(XY) - s_{\D}(X)\times s_{\D}(Y)$.
\end{definition}

If supports are unnormalized, extra factors $n$ are necessary.
In case of independence of both sides of a partial
implication $X\to Y$, we would have
$s(XY) = s(X)s(Y)$; therefore, both lift and leverage are
measuring deviation from independence: lift 
is the multiplicative deviation, whereas leverage 
measures it rather as an additive distance instead. 
Leverage was introduced in \cite{PSG} and,
under the name ``Novelty'', in \cite{LavracFlachZupan},
and received much attention via the Magnum Opus 
associator \cite{WebbMO}.
We find lift in the references going by 
several different names: it has been 
called {\em interest} \cite{BrinMS} or, 
in a slightly different but fully equivalent form, 
{\em strength} \cite{ShahLaksRS}; 
\emph{lift} seems to be catching up as a short name, 
possibly aided by the fact that the Intelligent Miner 
system from IBM employed that name. 
These notions allow us to exemplify that we are modifying
the semantics of our expressions: if we define the meaning of 
$X\to Y$ through confidence, then partial implications of the form 
$X\to Y$ and $X\to XY$ are 
always equivalent,
whereas, if we use lift, then they may not
be.
Note that, in case $X=\emptyset$, the lift
trivializes to~1. 
Also, if we are to use lift, 
then we must be careful to keep the right-hand side
$Y$ disjoint from the left-hand side: $X\cap Y = \emptyset$.

A related notion is:

\begin{definition}
\label{df:relconf}
\cite{LavracFlachZupan} 
The {\em relative confidence} of partial implication $X\to Y$,
also called {\em centered confidence} or {\em relative accuracy},
is $r_{\D}(X\to Y) = c_{\D}(X\to Y) - c_{\D}(\emptyset\to Y)$.
\end{definition}

Therefore, the relative confidence is measuring additively
the effect, on the support of the consequent $Y$,
of ``adding the condition'' or antecedent $X$.
Since $c_{\D}(\emptyset\to Y) = s_{\D}(Y)$, 
lift can be seen as comparing $c_{\D}(X\to Y)$ with
$c_{\D}(\emptyset\to Y)$, that is, effecting the same
comparison but multiplicatively this time:
$\ell(\hbox{$X\to Y$}) = \frac{s(XY)}{s(X)\times s(Y)} = \frac{c(X\to Y)}{s(Y)} = 
\frac{c(X\to Y)}{c(\emptyset\to Y)}$. 
Also, it is easy to check that leverage 
can be rewritten as 
$\l_{\D}(X\to Y) = s_{\D}(X)\times r_{\D}(X\to Y)$
and is therefore called also 
{\em weighted relative accuracy}~\cite{LavracFlachZupan}. 
Relative confidence has the potential to solve the 
``negative correlation'' 
objection to confidence, and all subsequent measures 
to be described here inherit this property as well.


An objection of a different sort is that lift and
leverage are symmetric. As the implicational syntax is asymmetric, 
they do not fit very well the directional intuition of
an expression like $X\to Y$; that is one of the reasons
behind the exploration of many other options.
However, to date,
none of the more sophisticate attempts seems to have gained a 
really noticeable ``market share''.
Most common implementations either offer a long list of
options of measures for the user to choose from 
(like \cite{Borgelt} for one), or employ the simpler
notions of confidence, support, lift, or leverage (for instance, 
Magnum Opus
\cite{WebbMO}). We believe that one must keep close 
to confidence and to deviation from independence. Confidence 
is the most natural option for many educated domain experts 
not specialized in data mining, and it provides actually 
a directionality to our partial implications. 


The vast majority of these alternatives
attempt at defining the quality of
partial implication $X\to Y$ relying only on the supports of $X$, $Y$, 
$XY$, or their complements. 
One major exception
is \emph{improvement} \cite{BAG},
which is the added confidence obtained by using the
given antecedent as opposed to any properly smaller one.
We discuss it and two other related quantities next. 
They are motivated again by our \emph{central intuition}:
if the confidence of a partial implication with a smaller 
antecedent and the same consequent is sufficiently high, 
the larger partial implication should not be provided in the output.
They have in common that their computation requires 
exploration of a larger space, however; we return to
this point in the next section.

\subsection{Improvement: Additive and Multiplicative}

The key observation for this section is that $X\to Y$ and $Z\to Y$,
for $Z\subset X$, provide different, independent information. 
From the perspective of confidence, either may have it arbitrarily 
higher than the other. For inequality in one direction, suppose 
that almost all transactions with $X$ have $Y$, but they are just 
a small fraction of those supporting $Z$, which mostly lack $Y$;
conversely, $Y$ might hold for most transactions having $Z$, but 
the only transactions having all of $X$ can be those without $Y$.
In Example~\ref{ex:running}, one can see that 
$c(\emptyset\to BC) < c(A\to BC)$
whereas $c(\emptyset\to C) > c(B\to C)$.

This fact underlies the difficulty in choosing a proper confidence 
bound. Assume that there exists a mild correlation
giving, say, $c(Z\to A) = 2/3$. If the threshold is set
higher, of course this rule is not found; but 
an undesirable side effect may appear: there may be many ways of
choosing subsets of the support of $Z$, by enlarging it a bit, 
where $Y$ is frequent enough to pass the threshold. Thus, often, 
in practice, the algorithms
enlarge $Z$ into various supersets $X_i$ so that all
the confidences $c(X_i\to A)$ do pass, and then $Z\to A$
is not seen, but generates dozens of very similar ``noisy'' rules, to be
manually explored and filtered. Finding the appropriate threshold
becomes difficult, also because, for different partial implications, 
this sort of phenomenon may appear
at several threshold values simultaneously.

Relative confidence tests confidence by a comparison to
what happens if the antecedent is replaced by 
one of its subsets in particular, namely $\emptyset$.
Improvement generalizes it by considering
not only the alternative partial implication $\emptyset\to Y$ but all 
proper subsets of the antecedent, as alternative antecedents,
and in the same additive form:

\begin{definition}
\label{df:improvement}
The {\em improvement} $X\to Y$,
where 
$X\neq\emptyset$, is
$i(X\to Y) = \min \{ c(X\to Y) - c(Z\to Y) \st Z\subset X \}$.
\end{definition}

The definition is due to \cite{BAG}, where only association rules
are considered,
that is, cases where $|Y|=1$.
The work on productive rules~\cite{Webb07} is related: these
coincide with the rules of \emph{positive improvement}. 
In~\cite{LiuHsuMa}, improvement 
is combined with further pruning on the basis of the $\chi^2$ value.
We literally quote from \cite{BAG}:
``A rule with negative improvement is
typically undesirable because the rule can be simplified to yield a proper sub-rule that is
more predictive, and applies to an equal or larger population due to the antecedent containment 
relationship. An improvement greater than 0 is thus a desirable constraint in almost
any application of association rule mining. A larger minimum on improvement is also often
justified because most rules in dense data-sets are not useful due to conditions or combinations 
of conditions that add only a marginal increase in confidence.''

The same process, and with the same intuitive justification, 
can be applied to lift, which is, actually,
a multiplicative, instead of additive, version
of relative confidence as indicated above:
$\ell(X\to Y) = c(X\to Y) / c(\emptyset\to Y)$.
Taking inspiration in this
correspondence, we 
studied
in~\cite{BalDog} a multiplicative 
variant of improvement that generalizes lift, exactly in 
the same way as improvement generalizes relative confidence:

\begin{definition}
\label{df:multimpr}
The {\em multiplicative improvement} of $X\to Y$,
where $X\neq\emptyset$, is
$m(X\to Y) = \min \{ c(X\to Y)/c(Z\to Y) | Z\subset X \}$.
\end{definition}

In Example~\ref{ex:running}, the facts that 
$c(A\to BC) = 4/5$ 
and 
$c(\emptyset\to BC) = 3/4$ 
lead to 
$i(A\to BC) = 4/5 - 3/4 = 0.05$ and 
$m(A\to BC) = (4/5) / (3/4) \approx 1.066$.
Here, as the size of the antecedent is 1,
there is one single candidate $Z = \emptyset$ 
to proper subset of the antecedent and,
therefore, improvement coincides with relative
confidence, and multiplicative improvement 
coincides with lift. For larger left-hand 
sides, the values will be different in general.

\subsection{Rule Blocking}

Attempting at formalizing the same part of the
\emph {central intuition}, we proposed in \cite{Bal09} a notion 
of ``rule blocking'', 
where a
smaller antecedent $Z\subset X$ would ``block''
(that is, suggest to omit)
a given partial implication $X\to Y$. 
We will compare the number of tuples having $XY$ 
(that is, having $Y$ within the supporting set of $X$)
with the quantity that would be predicted from the confidence 
of the partial implication $Z\implies Y$, that applies to a larger
supporting set: we are going to bound the relative
error incurred if the support $s(X)$ and
the confidence of $Z\to Y$ are employed to
approximate the confidence of $X\to Y$.

More precisely, let $c(Z\implies ZY) = c$. If $Y$ is
distributed along the support of $X$ at the same ratio as along 
the larger support of $Z$, we would expect $s(XY)\approx c \times s(X)$:
we consider the relative error committed by
$c\times s(X)$ used as an approximation to $s(XY)$
and, if the error is low, we consider that $Z\to Y$ 
is sufficient information about $X\to Y$ and dispose of
this last one.

\begin{definition}
\label{df:blocking}
\cite{Bal09} 
$Z\subset X$ blocks $X\to Y$ at blocking threshold $\epsilon$ when
$$
\frac{s(XY) - c(Z\implies Y)s(X)}{c(Z\implies Y)s(X)} \leq \epsilon.
$$
\end{definition}


In case the difference in the numerator is negative, it would mean 
that $s(XY)$ is even lower than what $Z\implies Y$ would suggest. 
If it is positive but the quotient is low, $c(Z\implies Y)\times s(X)$
still suggests a good approximation to $c(X\implies Y)$, and
the larger partial implication $X\implies Y$
does not bring high enough confidence to be considered
besides $Z\implies Y$, a simpler one: it remains blocked. 
But, if the quotient is larger, 
and this happens for all $Z$, then $X\implies Y$ becomes interesting 
since its confidence is higher enough than suggested by other partial implications of 
the form $Z\implies Y$ for smaller antecedents $Z$.
Of course, the higher the block threshold, the more demanding 
the constraint is. 
Note that, in the presence of a support threshold $\tau$,
$s(ZY) \geq s(XY) > \tau$ or a similar inequality would be
additionally required.
The value $\epsilon$ is intended to take positive but
small values, say around 0.2 or lower.
In Example~\ref{ex:running}, $\emptyset$ blocks $A\to BC$ 
at blocking threshold $1/15 \approx 0.066$.

Rule blocking 
relates to multiplicative improvement as follows:

\begin{proposition}
\label{pr:block}
The smallest blocking threshold at which $X\to Y$ is
blocked is $m(X\to Y)-1$.
\end{proposition}

\begin{proof}
As everything around is finite, this is equivalent to proving
that $Z\subset X$ blocks $X\to Y$ at block 
threshold $\epsilon$
if and only if $\frac{c(X\to Y)}{c(Z\implies Y)} -1 \leq\epsilon$,
for all such $Z$.
Starting from the definition of blocking, 
multiplying both sides of the inequality
by $c(Z\to Y)$, separating the two
terms of the left-hand side, replacing $s(XY)/s(X)$ by its
meaning, $c(X\to Y)$, and then solving first for $c(Z\to Y)$
and finally for $\epsilon$, we find the stated equivalence.
All the algebraic manipulations are reversible.
\end{proof}
 
\subsection{Ongoing: Conditional Weighted Versions of Lift and Leverage}

We propose here one additional step to enhance the flexibility 
of both lift and leverage by
considering their action, on the same partial implication, but with respect to
many different subsets of the dataset, and under a weighting
scheme that leads to different existing measures according to
the weights chosen.

For a given partial implication $X\to Y$, we consider many limited
views of the dataset, namely, all its projections into subsets
of the antecedent. We propose to measure a weighted variant
of the lift and/or the leverage of the same partial implication in all 
these projections, and evaluate as the quality of the partial implication 
the minimum value thus obtained. That is, we want our 
high-quality partial implications not only to have high lift or leverage, 
but also to maintain it when we consider projections of 
the dataset on the subsets of the antecedent. We call the 
measures obtained \emph{conditional weighted} lift and leverage.

\begin{definition}
\label{df:cwliftleverage}
Assume $X\cap Y = \emptyset$.
Let $w$ be a weighting function associating a
weight (either a positive real number or $\infty$)
to each proper subset of $X$.
The {\em conditional weighted lift} of partial implication $X\to Y$ is 
$\ell'_{\D,w}(X\to Y) = \min\{ w(Z)\ell_{\D_Z}(\hbox{$X\to Y$}) \st Z\subseteq X \}$.
The {\em conditional weighted leverage} 
of partial implication $X\to Y$ is 
$\l'_{\D,w}(X\to Y) = \min\{ w(Z)g_{\D_Z}(X\to Y) \st Z\subseteq X \}$.
\end{definition}

These notions can be connected to other existing notions
with unificatory effects. We only state here one such connection.
Further development will be provided in a future paper in preparation.

\begin{proposition} 
For inverse confidence weights, conditional weighted leverage is 
improvement: for all $X\to Y$, $\l'_{\D,w}(X\to Y) = i(X\to Y)$
holds for the weighting function $w_r(Z) = c_{\D}(Z\to X)^{-1}$.
\end{proposition}

\section{Support Ratio and Confidence Boost}

From the perspective of our \emph{central intuition},
the previous section has developed, essentially issues
related to smallish antecedents. This is fully appropriate
for the discussion of association rules, which were defined
originally as partial implications with singleton consequents.
We now briefly concentrate on largish consequents, and then 
join both perspectives.

\subsection{Support ratio}

The support ratio was employed first, to our knowledge, in
\cite{KryszIDA}, where no particular name was assigned to it.
Together with other similar quotients, it was introduced 
in order to help obtaining faster algorithmics.

\begin{definition}
\label{d:suppratio}
In the presence of a support threshold $\tau$, 
the {\em support ratio} of a partial implication 
$X\to Y$ 
is 
$$
\sigma(X\to Y)
 = 
\frac{s(XY)}{\max \{ s(Z) | XY\subset Z, \, s(Z) > \tau \}}.
$$
\end{definition}

We see that this quantity depends 
on $XY$ but not 
on the antecedent~$X$ itself.
In Example~\ref{ex:running}, we find that 
$\sigma(A\to BC) = 4/3$.

\subsection{Confidence Boost}
\label{ss:cboost}

\begin{definition}
\label{d:boost}
The {\em confidence boost} of a partial implication $X\to Y$ 
(always with $X\cap Y = \emptyset$) is 
$\beta(X\to Y) = {}$
$$ 
 \frac{c(X\to XY)}{\max \{ c(X'\to X'Y') \st
 (X\to XY) \not\equiv (X'\to X'Y'), \, X'\subseteq X, \, Y\subseteq Y'\}}.
$$
where the partial implications in the denominator are implicitly 
required to clear the support threshold, in case 
one is enforced:
$s(X'\to X'Y') > \tau$.
\end{definition}

Let us explain the interpretation of this parameter.
Suppose that $\beta(X\to Y)$ is low, say
$\beta(X\to Y)\leq b$, where $b$ is just slightly larger than~1. 
Then,
according to the definition, there must exist 
some {\em different} partial implication $X'\to X'Y'$, with $X'\subseteq X$ and 
$Y\subseteq X'Y'$, such that $\frac{c(X\to Y)}{c(X'\to Y')}\leq b$, 
or $c(X'\to Y')\geq c(X\to Y)/b$. 
This inequality says that the partial implication $X'\to Y'$, stating that transactions 
with $X'$ tend to have $X'Y'$, has a confidence relatively high, not much 
lower than that of $X\to Y$; equivalently, the confidence of $X\to Y$ 
is not much higher (it could be lower) than that of $X'\to Y'$. But 
all transactions having $X$ do have $X'$, and all transactions having $Y'$ 
have $Y$, so that the confidence found for $X\to Y$ is not really that 
novel, given that it does not give so much additional confidence over a
partial implication that states such a similarly confident, and intuitively stronger, 
fact, namely $X'\to Y'$. 

This author has developed a quite successful open-source partial 
implication miner based on confidence boost ({\tt yacaree.sf.net});
all readers are welcome to experiment with it and provide feedback.
We note also that 
the confidence width alluded to in Section~\ref{ss:cwidth},
while having different theoretical and practical properties,
is surprisingly close in definition to confidence boost.
See \cite{BalTKDD} for further discussion of all these issues.
Confidence boost fits the general picture as follows:

\begin{proposition}
$\beta(X\to Y) = \min\{ \sigma(X\to Y), m(X\to Y) \}$.
\end{proposition}

The inequalities $\beta(X\to Y)\leq \sigma(X\to Y)$
(due to \cite{BalTirZor10a}) and 
$\beta(X\to Y)\leq m(X\to Y)$ are simple to argue:
the consequent leading to the support ratio, or the
antecedent leading to the multiplicative improvement,
take a role in the denominator of confidence boost.
Conversely, taking the maximizing partial implication
in the denominator, if it has the same antecedent $X$
then one obtains a bound on the support ratio whereas,
if the antecedent is properly smaller, a bound on
the multiplicative improvement follows.

In Example~\ref{ex:running}, since $\sigma(A\to BC) = 4/3$
and $m(A\to BC) = (4/5) / (3/4)$, which is smaller, we obtain
$\beta(A\to BC) = (4/5) / (3/4) \approx 1.066$.

A related proposal in \cite{KryszPKDD} suggests to
minimize directly the antecedents and maximizing the
consequents, within the confidence bound, and in a 
context where antecedents and consequents are kept 
disjoint. This is similar to statement $(3)$ in
Lemma~\ref{l:redchar}, except that, there, one
maximizes jointly consequent and antecedent. If
consequents are maximized separately, then the
\emph{central intuition} fails, but there is an
interesting connection with confidence boost;
see~\cite{BalTKDD}.

The measures in this family of improvement, including
conditional weighted variants and also confidence boost,
tend to require exploration of larger spaces of antecedents
compared to simpler rule quality measures. This objection
turns out not to be too relevant because human-readable 
partial implications have often just a few items in the 
antecedent. Nontrivial algorithmic proposals for handling 
this issue appear as well in \cite{BalTKDD}.

\subsection{Ongoing Developments}

We briefly mention here the following observations. 
First, like in Section~\ref{ss:bstar}, a variant 
of confidence boost appropriate for closure-based analysis
exists \cite{BalTKDD}. Second, both variants
trivialize if they are applied directly, in their literal
terms, to full implications. However, the intuitions leading
to confidence boost can be applied as well to full implications.
In future work, currently in preparation, we will discuss
proposals for formalizing the same intuition in the context
of full implications.

\section{Evaluation of Evaluation Measures}

We have covered just a small fraction of the evaluation
measures proposed to endow with useful semantics the partial
implication connective. All of those attempt, actually, at
capturing a potential (but maybe nonexisting) ``na\"\i{}ve
concept'' of interesting partial implication from the 
perspective of an end user. Eventually, we would like to
find one such semantics that fits as best as possible that
hypothetical na\"\i{}ve concept.

We can see no choice but to embark, at some point, in the
creation of resources where, for specific datasets, the
interest of particular implications is recorded as per
the assessment of individual humans. Some approximations 
to this plan are Section~5.2 of \cite{BalTKDD},
where the author, as a scientific expert, subjectively
evaluates partial implications obtained from abstracts or
scientific papers; a similar approach in \cite{MINI} using
PKDD abstracts; and the work in \cite{BalTirZor10b,ZorGB}
where partial implications found on educational datasets
from university course logs are evaluated by the teachers 
of the corresponding courses. These preliminary experiments
are positive and we hope that a more ambitious attempt
could be made in the future along these lines.

The idea of evaluating associators through
the predictive capabilities of the rules found has
been put forward in several sources, e.g.~\cite{MHF}. 
The usage of
association rules for direct prediction (where the 
``class'' attribute is forced to occur in the consequent)
has been widely studied (e.g.~\cite{YinHan}). 
In \cite{MHF}, two 
different associators are employed to find rules
with the ``class'' as consequent,
and they are compared in terms of predictive accuracy.
This scheme is inappropriate to evaluate 
our proposals for the semantics of
partial implications, because, first,
we must focus on single pairs of attribute and
value as right-hand side, thus making it useless to consider larger
right-hand sides;
and, also, the classification
will only be sensible to minimal left-hand sides independently of
their confidences.

In \cite{BalDog}, we have deployed
an alternative framework that allows us to evaluate the 
diverse options of semantics for association rules,
in terms of their usefulnes for subsequent predictive tasks.
By means of a mechanism akin to the AUC measure for
predictor evaluation, 
we have focused on potential accuracy improvements
of predictors on given, public, 
standard benchmark datasets, if one
more Boolean column is added, namely, one that is true exactly for
those observations that are exceptions to one association
rule: the antecedent holds but the consequent does not.
In a sense, we use the association rule as a ``hint of
outliers'', but, instead of removing them, we simply offer
direct access to this label to the predictor, through the
extra column.
Of course, in general this may lead astray the predictor 
instead of helping it. 
Our experiments suggest that leverage, support, 
and multiplicative improvement tend to be better than
the other measures with respect to this evaluation score.

\subsection{Ongoing Developments}

We are currently developing yet new frameworks that,
hopefully, might be helpful in assessing the relative merits
of the different candidates for semantics of partial implications,
put forward often as rule quality measures. One of them 
resorts to an empirical application of approximations to
the MDL principle along the lines of Krimp \cite{krimp}.
A second idea is to make explicit the dependence on
alternative partial implications, in the sense that $X\to Y$
would mean, intuitively, that $Y$ appears often on the
support of $X$ and that, barring the presence of some
other partial implication to the contrary, it is 
approximately uniformly distributed there. These avenues
will be hopefully explored along the coming months or years.
A common thread is that additional statistical knowledge,
along the lines of the self-sufficient itemsets of 
Webb~\cite{Webb10}, for instance, is expected to be at play 
in the future developments of the issue of endowing the
partial implication connective with the right intuitive
semantics.

\bibliographystyle{abbrv}

\end{document}